\documentclass[12pt]{iopart} % JPG
\usepackage{graphicx}
\usepackage{comment}
\usepackage{hyperref}
\usepackage{amssymb,bm,calc}

\begin{document}

\title{Charged current quasi elastic scattering of muon neutrino with nuclei}

\author{Kapil Saraswat $^{1}$ ,  Prashant Shukla$^{2,3,\ast}$, Vineet Kumar$^{2}$, Venktesh Singh$^{1}$}

\address{$^{1}$ Department of Physics, Institute of Science, Banaras Hindu University, 
Varanasi 221005, India.}
\address{$^{2}$ Nuclear Physics Division, Bhabha Atomic Research Center, Mumbai 400085, India}
\address{$^{3}$ Homi Bhabha National Institute, Anushakti Nagar, Mumbai 400094, India.}
\ead{$^{\ast}$ pshuklabarc@gmail.com}

\date{\today}

\begin{abstract}
We present a study on the charge current quasi elastic scattering of $\nu_\mu$ from
nucleon and nuclei which gives a charged muon in the final state. To describe 
nuclei, the Fermi Gas model has been used with proposed Pauli suppression factor. 
The diffuseness parameter of the Fermi distribution has been obtained using 
experimental data.
We also investigate different parametrizations for electric and magnetic Sach's form 
factors of nucleons. Calculations have been made for CCQES total and differential cross 
sections for the cases of $\nu_{\mu}-N$, $\nu_{\mu}-^{12}C$ and $\nu_{\mu}-^{56}Fe$ scatterings
and are compared with the data for different values of the axial mass. The present model 
gives excellent description of measured differential cross section for all the systems.
%The Pauli suppression reduce the cross section by 10 \% even at higher 
%neutrino energy above 1 GeV. 
\vspace{0.6in}  
\end{abstract}


\maketitle

\section{Introduction}
The neutrinos produced in the upper atmosphere due to the bombardment of cosmic rays are 
one of the best tools for the study of neutrino oscillations. There are many ongoing and 
proposed experiments worldwide to study the phenomena of neutrino oscillations  
\cite{Ashie:2005ik, Takeuchi:2011aa, Aliu:2004sq, Ahn:2002up, Ahn:2006zza, Ahmed:2015jtv}. 
The neutrinos with energies between 1 to 3 GeV form bulk of the 
signal in the detector. The neutrino in this energy range can interact with matter 
by many processes such as quasi elastic scattering, interaction via resonance 
pion production, and deep inelastic scattering~\cite{Formaggio:2013kya}.
There is also coherent pion production process in neutrino nucleus scattering~\cite{Saraswat:2016kln}.
The charge current interactions of neutrinos inside the 
matter are important since the experiments measure the recoil muon produced in such 
interactions. The materials like carbon, iron~\cite{Ahmed:2015jtv} and argon offer a 
convenient detector media. We present the studies with carbon and iron. 

  At the lowest neutrino energies, the interactions with nucleon are either elastic 
or quasi elastic in which the nucleon recoils intact. In the neutral current (NC) 
elastic scattering, all neutrinos and all anti-neutrinos can scatter off both neutrons 
and protons: $\nu + N \rightarrow \nu + N$. When neutrinos acquire sufficient energy, 
they can also undergo the charged current interactions: 
$\nu_{l} + n \rightarrow l^{-} + p$ and $\bar \nu_{l} + p \rightarrow l^{+} + n$. 
This is called charged current quasi-elastic scattering (CCQES) as a charged lepton 
mass is created. CCQES interactions are important to neutrino physics for two reasons. 
First, the nucleon recoils intact that enables to measure weak nucleon form factors 
which are difficult to measure in other scattering probes. Second, the two body 
interaction enables the kinematics to be completely reconstructed (if one ignores 
two-nucleon contribution) and hence the 
initial neutrino energy can be determined which is crucial for measuring the 
oscillation parameters. Thus, a reliable description of the neutrino quasi elastic 
scattering (QES) processes (particularly on nuclear targets) is essential for precision 
studies of neutrino oscillation parameters, such as mass splitting and mixing angles 
\cite{Fukuda:2000np, Michael:2006rx, Adamson:2009ju, Adamson:2011fa, Adamson:2016tbq, Adamson:2016xxw, Adamson:2017qqn}.

The most popular model for the CCQES calculations is the Llewellyn Smith (LS) 
model~\cite{LlewellynSmith:1971zm}. The Fermi Gas Model~\cite{Smith:1972xh} with 
Pauli suppression condition is used in the case of bound nucleon to include nuclear 
modifications. These models have advantage that they can be readily incorporated into 
existing neutrino Monte Carlo generators \cite{Andreopoulos:2009rq} 
\cite{Gallagher:2002sf, Hayato:2002sd, Casper:2002sd}, though there exist 
more sophisticated calculations of quasi elastic scattering such as relativistic 
distorted-wave impulse approximation \cite{Butkevich:2010cr}. 
A global analysis \cite{Kuzmin:2007kr} of the QES cross sections measured in high 
energy $\nu_{\mu}$ experiments on nuclear targets finds a value of axial mass 
$M_{A}$ = 0.977 $\pm$ 0.016 GeV. There are recent measurements of differential 
and total cross sections for QES of $\nu_{\mu},\bar{\nu}_\mu$ of energies above 4 GeV 
on carbon by the NOMAD collaboration \cite{Lyubushkin:2008pe}. The NOMAD analysis of 
cross sections assuming free nucleon model yields a value of 
$M_A$ = $1.05\pm0.02\pm$0.06~GeV. 
The nuclear effects as will be shown here reduce the 
cross section by 10 \% even at higher neutrino energy above 1 GeV which implies 
that a larger fit value of $M_A$ is expected if the nuclear effects are included.
  The QES cross sections for low energy ($\approx$) 1 GeV neutrino scattering on 
carbon measured by the MiniBooNE detector 
\cite{AguilarArevalo:2007it, AguilarArevalo:2011sz, Juszczak:2010ve, AguilarArevalo:2010cx}
is 20$\%$ larger than the model calculations. The model calculations of MiniBooNE 
analysis use a large value of axial mass (1.35 GeV) which increases the total cross 
section at all energies along with an empirical parameter 
in the Pauli Blocking condition which decreases the cross section at lower energies  
 \cite{AguilarArevalo:2010zc}. The data from K2K with oxygen 
target \cite{Gran:2006jn} also require a higher value $M_A$ = $1.2 \pm 0.12$ GeV.

 To get a better description of the cross-section data at low energy, the work in 
Ref.~\cite{Kolupaeva:2016bfg} parametrizes $M_A$ as a function of energy which 
results in a higher value of $M_A$ at low energy.
  A recent work uses an axial vector form factor obtained from a generalized axial 
vector meson dominance model with prior uncertainty band \cite{Amaro:2015lga}. 
A relativistic Fermi gas model is then used to describe the MiniBooNE data in 
neutrino energy range from 0.5 to 1.5 GeV within uncertainties.
  A superscaling approach (SuSA) based on the analysis of electron nucleon scattering 
improved with relativistic mean field theory effect is used in Ref.~\cite{Megias:2016fjk}.
 They also study the contribution of two nucleon knockout reaction in neutrino
nucleus interaction.
The work in Refs.~\cite{Amaro:2010sd, Amaro:2011aa} used the empirical SuSA 
scaling function to describe the CCQE MinibooNE data including 2p2h (two nucleons producing 
two holes) contributions. 
 The calculations underestimates the data even after adding the 2p2h contibutions.
In the present work we do not include the 2p2h contribution.
%%%%%%%%%%%%%%%%%%%%%%%%%%%%%%%%%%%%%%%%%%%%%%%%%%%%%%%%%%%%%%%%%%%%%%%%%%%%%%%%%%%%%%%%%%%%%

  In this work, we calculate the neutrino-nucleon CCQES cross section using 
Llewellyn Smith model. We also investigate different parametrizations for electric 
and magnetic Sach's form factors of nucleons. For the cases of nuclei, the Fermi 
Gas model has been used with Pauli blocking.
%The effective binding energy of nucleons 
%inside the nuclei has been used as a parameter to get a good description of the data.
  Calculations have been made for CCQES total and differential cross sections
for the case of $\nu_{\mu}-N$, $\nu_{\mu}-^{12}C$ and $\nu_{\mu}-^{56}Fe$ scattering and 
are compared with the data with the aim of obtaining appropriate value of
the axial mass. 
%%%%%%%%%%%%%%%%%%%%%%%%%%%%%%%%%%%%%%%%%%%%%%%%%%%%%%%%%%%%%%%%%%%%%%%%%%%%%%%%%%%%%%%%%%%%%
%%%%%%%%%%%%%%%%%%%%%%%%%%%%%%%%%%%%%%%%%%%%%%%%%%%%%%%%%%%%%%%%%%%%%%%%%%%%%%%%%%%%%%%%%%%%%

\section{The Model of Neutrino-Nucleon Quasi Elastic Scattering}

 The charged current quasi elastic (CCQES) neutrino nucleon differential 
cross section for a nucleon in the rest is given by \cite{LlewellynSmith:1971zm}
\begin{eqnarray}
\frac{d\sigma^{free}}{dQ^{2}} = \frac{M_{N}^2~G_{F}^2~\cos^{2}\theta_{c}}
{8\pi E_{\nu}^{2}}
\Bigg[A(Q^{2}) \pm~\frac{B(Q^{2})~(s-u)}{M_{N}^2} 
 + \frac{C(Q^{2})~(s-u)^{2}}{M^{4}_{N}} \Bigg].
\label{smith1971first}
\end{eqnarray}
Here, $M_{N}$ is the mass of nucleon, $G_{F}$ (=1.16 $\times$ 10$^{-5}$
GeV$^{-2})$ is the Fermi coupling constant and $\cos\theta_{c}$(=0.97425) 
is the Cabibbo angle. In terms of the mandelstam 
variables $s$ and $u$, the relation $s-u=4M_{N}E_{\nu}-Q^{2}-m_{l}^{2}$,
where $m_{l}$ is the mass of muon, $E_{\nu}$ is the neutrino energy and $Q^{2}$ 
is the square of the momentum transfer from neutrino to outgoing muon.

The functions $A$, $B$  and $C$ can be written in the following form 
\cite{LlewellynSmith:1971zm}
\begin{eqnarray}
A(Q^{2}) &=& \frac{(m_{l}^{2}+Q^2)}{M^2_{N}}\Bigg \{\Bigg[(1+\tau)F_{A}^{2}-(1-\tau)
(F^{V}_{1})^{2}  \nonumber \\
&~& +\tau(1-\tau)(F^{V}_{2})^{2}  + 4\tau~F^{V}_{1}F^{V}_{2} \Bigg]  \nonumber \\
&~& - \frac{m_{l}^{2}}{4M^{2}_{N}}\Bigg[(F^{V}_{1}+F^{V}_{2})^2+(F_{A}+2F_{P})^{2} 
 - 4(1+\tau) F_{P}^{2}  \Bigg] \Bigg \},       \\
B(Q^2) &=& \frac{Q^2}{M^2_{N}}~F_{A}~(F^{V}_{1}+F^{V}_{2})~, \\
C(Q^{2}) &=& \frac{1}{4}~\Big[F_{A}^{2}+(F^{V}_{1})^{2}+\tau (F^{V}_{2})^{2} \Big]~.
\end{eqnarray}
\noindent
Here, $\tau = Q^2/(4 M^2)$. The form factors used for neutrino and antineutrino 
scatterings are the same because of the charge symmetry of the matrix element. 
The function $F_{A}$ is the axial form factor, $F_{P}$ is the pseudoscalar form factor 
and $F^{V}_{1}, F^{V}_{2}$ are the vector form factors. 

 The axial form factor $F_{A}$ can be written in the dipole 
form~\cite{Stoler:1993yk} as
\begin{equation}
F_{A}(Q^2)=\frac{g_{A}}{(1+\frac{Q^2}{M^{2}_{A}})^{2}},
\label{axialformequation}
\end{equation}
where $g_{A}$ (=~-1.267) is the axial vector constant and $M_{A}$ is the axial mass.
 The pseudoscalar form factor $F_{P}$ can be calculated from the axial form factor
 $F_{A}$~\cite{Bernard:2001rs} as
\begin{equation}
F_{P}(Q^2)=\frac{2~M^2_{N}}{Q^2+m^{2}_{\pi}}~F_{A}(Q^2)~,
\end{equation}
where $m_{\pi}$ is the mass of pion. 
  The vector form factors $F^{V}_{1}$ and $F^{V}_{2}$ can be written as
\cite{Stoler:1993yk} \cite{Budd:2003wb} 
\begin{eqnarray}
F^{V}_{1} &=& 
\frac{\Big[G^{p}_{E}(Q^2) - G^{n}_{E}(Q^2)\Big] + 
\tau \Big[G^{p}_{M}(Q^2) - G^{n}_{M}(Q^2)\Big]}{1 + \tau} ~,\\
F^{V}_{2} &=& 
\frac{\Big[G^{p}_{M}(Q^2) - G^{n}_{M}(Q^2)\Big] - 
\Big[G^{p}_{E}(Q^2) - G^{n}_{E}(Q^2)\Big]}{1 + \tau}  ~.
\end{eqnarray} 
 Here, $G^{p,n}_{E}$ and $G^{p,n}_{M}$ are respectively the electric and magnetic 
Sach's form factors
of nucleons (proton and neutron).
  There are many parametrizations of these form factors which are
obtained by fitting the electron scattering data and are given by
Galster~\cite{Galster:1971kv}, Budd et al.~\cite{Budd:2004bp}, Bradford
et al.~\cite{Bradford:2006yz}, 
Bosted~\cite{Bosted:1994tm} and Alberico et al.~\cite{Alberico:2008sz}. 
We use Galster parametrizations in our calculation. We find that all the parametrizations
produce almost same results. Such a comparison with Galster and the latest Alberico's 
parametrizations is presented in the results section.

  The electric and magnetic Sach's form factors given by Galster~\cite{Galster:1971kv}
are as follows  
\begin{eqnarray}
G^{p}_{E} &=& G_{D}(Q^2) ,       \nonumber \\
G^{p}_{M}(Q^2) &=& \mu_{p}~G_{D}(Q^2) ,  \nonumber \\
%G^{n}_{E} &=& 0,   \nonumber \\ 
G^{n}_{M}(Q^2) &=& \mu_{n}~G_{D}(Q^2) .
\end{eqnarray}
 For the electric form factor of neutron, we use the parametrization given by 
Krutov et. al.~\cite{Krutov:2002tp} as
\begin{equation}
G^{n}_{E}(Q^2)=~-\mu_{n}~\frac{0.942 ~\tau}{(1+ 4.61 ~\tau)}~G_{D}(Q^2).
\end{equation}
The magnetic moment of proton $\mu_{p}=2.793$ and that of neutron 
$\mu_{n}=~-1.913$. The dipole form factor $G_{D}(Q^2)$ is given by \cite{Stoler:1993yk} 
\begin{equation}
G_{D}(Q^2) = \frac{1}{\Big(1 + \frac{Q^{2}}{M^{2}_{v}} \Big)^{2}}~,
\end{equation}
with $M^{2}_{v}=$0.71~GeV$^{2}$. 

  The Sach's form factor given by Alberico et al.~\cite{Alberico:2008sz} are
\begin{eqnarray}
G^{p}_{E}(Q^{2}) &=& \frac{1 - 0.14 \tau}
{1 + 11.18 \tau + 15.18 \tau^{2} + 23.57 \tau^{3}}~, \nonumber \\
\frac{G^{p}_{M}(Q^{2})}{\mu_{p}} &=& \frac{1 + 1.07 \tau}
{1 + 12.30 \tau + 25.43 \tau^{2} + 30.39 \tau^{3}}~,  \nonumber \\
G^{n}_{E}(Q^{2}) &=& -\frac{0.10}{(1 + 2.83 Q^{2})^{2}} + 
\frac{0.10}{(1 + 0.43 Q^{2})^{2}}~,  \nonumber \\
\frac{G^{n}_{M}(Q^{2})}{\mu_{n}} &=& \frac{1 + 2.13 \tau}
{1 + 14.53 \tau + 22.76 \tau^{2} + 78.29 \tau^{3}}~.
\end{eqnarray}
  We study the effect of Galster and Alberico parametrizations of electromagnetic 
form factors on CCQES cross section of neutrino nucleon scattering.

The neutrino QES total cross section for a free nucleon is calculated 
as~\cite{amin:2006}
\begin{equation}
\sigma^{free}(E_{v})=~\int^{Q^{2}_{max}}_{Q^{2}_{min}}~dQ^{2}~
\frac{d\sigma^{free}(E_{\nu},Q^{2})}{dQ^{2}}~,
\end{equation}
where
\begin{eqnarray}
 Q^{2}_{min} &=&~ -m^{2}_{l} + 2~E_{\nu}~(E_{l}-|\vec k^{'}|)    \nonumber \\
 &=&~ \frac{2E^{2}_{\nu}M_{N}-M_{N}~m^{2}_{l}-E_{\nu}m^{2}_{l} - E_{Q}}{2E_{\nu}+M_{N}},
\label{qsquareminres}
\end{eqnarray}
\begin{eqnarray}
 Q^{2}_{max} &=&~ -m^{2}_{l} + 2~E_{\nu}~(E_{l}+ |\vec k^{'}|)    \nonumber \\
 &=&~ \frac{2E^{2}_{\nu}M_{N}-M_{N}~m^{2}_{l}+E_{\nu}m^{2}_{l}+E_{Q}}{2E_{\nu}+M_{N}},
\label{qsquaremaxres}
\end{eqnarray}
where $E_{Q} = E_{\nu}\sqrt{(s-m^{2}_{l})^{2}-2(s+m^{2}_{l})M^{2}_{N}+M^{4}_{N}}$, 
$s=M^{2}_{N}+2M_{N}E_{\nu}$ and  $E_{l}$ and $\vec k^{'}$ are the energy and momentum 
of the charged lepton.

{\bf The Fermi Gas Model}

  The cross section for the neutrino scattering with a nucleon in a nucleus 
is smaller than that in the case of  free nucleon scattering.
 The nucleus can be treated in terms of the Fermi gas model where the nucleons
move independently (Fermi motion) within the nuclear volume in a constant 
binding potential generated by all nucleons.  In the Fermi gas model, all the states 
up to the Fermi momentum $k_F$ are occupied. The Pauli blocking implies that the cross 
section for all the interactions leading to a final state nucleon with a momentum smaller 
than $k_{F}$ is equal to zero. There are many prescriptions for the Fermi model 
\cite{LlewellynSmith:1971zm, Smith:1972xh, Kuzmin:2007kr} and Pauli blocking.

 The differential cross section per neutron for the charged current neutrino-nucleus 
quasi elastic scattering is given by 
\begin{eqnarray}
\frac{d\sigma^{nucleus}(E_{\nu})}{dQ^{2}} &=& \frac{2 V }{(A - Z) (2 \pi)^{3}}
\int^{\infty}_{0} 2\pi k_n^2 dk_n d(\cos\theta) f(\vec{k_{n}}) \nonumber \\ 
&~& ~S(\nu - \nu_{min}) \frac{d\sigma^{free}(E^{eff}_{\nu}(E_{\nu},\vec{k_{n}}))}
{dQ^2}~. 
\label{qesdifferentialcrosssection}
\end{eqnarray}
Here, the factor 2 accounts for the spin of the neutron, 
$2 V/((A - Z) (2 \pi)^{3}) = \rho_{0}$ and $V$ is the volume of the 
the nucleus. $d\sigma^{free}/dQ^{2}$ is the neutrino QES differential 
cross section for free neutron at rest given by Eq.~\ref{smith1971first}. 
The effective neutrino energy $E^{eff}_{\nu}$ in the 
presence of the Fermi motion of the nucleon is given by 
\begin{equation}
E^{eff}_{\nu}=\frac{(s^{eff}-M^{2}_{n})}{2M_{n}}~.
\end{equation}
Here, $s^{eff}=M^{2}_{n}+2 E_{\nu} \Big(E_{n}-k_{n}~\cos(\theta) \Big)~$
and the neutron energy is $E_{n} = \sqrt{k^{2}_{n} + M^{2}_{n}}$ in terms of 
neutron momentum $k_{n}$ and mass $M_{n}$. 
The fermi distribution for non-interacting particles at zero temperature 
sharply drops at fermi momentum. In reality, the fermi distribution 
would drop smoothly with a diffuseness $a = kT$.  
The Fermi distribution function $f(\vec k_{n})$ is defined as 
\begin{equation}
f(k_{n}) = \frac{1}{1 + \exp(\frac{k_{n} - k_{F}}{a})}.
\end{equation}
Here, $k_{F}$ is the mean Fermi momentum. 
The normalization $\rho_{0}$ is given as 
\begin{equation}
\rho_{0} = \frac{1}{\frac{4\pi}{3} k^{3}_{F} \Big( 1 + \frac{\pi^{2} a^{2}}{k^{2}_{F}}\Big)}.
\end{equation}
The Pauli suppression factor is given by 
\begin{equation}
S(\nu - \nu_{min}) = \frac{1}{1 + \exp(-\frac{(\nu - \nu_{min})}{a})}.
\end{equation}
Here, the variable $\nu = (Q^{2} + M^{2}_{p} - M^{2}_{n})/(2 M_{n})$ is the energy 
transfer in the collisions and $\nu_{min}$ is obtained by Pauli Blocking and 
binding energy ($E_{B}$) considerations:
$\nu_{min} = \sqrt{k^{2}_{F} + M^{2}_{p}} - \sqrt{k^{2}_{n} + M^{2}_{n}} + E_{B}$. 
The value of binding energy is 10 MeV for both carbon and iron nuclei.
The $Q^{2}$ is restricted in $Q^{2}_{min} ~\textless ~ Q^{2} ~ \textless ~ Q^{2}_{max}$ 
where  $Q^{2}_{min}$ and $Q^{2}_{max}$ are calculated by 
Eq.~\ref{qsquareminres} and \ref{qsquaremaxres} but with $E^{eff}_{\nu}$.

  The value of the Fermi momentum ($k_{F}$) for the carbon nucleus is taken 
as 0.221 GeV from Ref.~\cite{Moniz:1971mt} . For the iron nucleus, $k_{F}$ for both the 
neutron and proton is taken as 0.260 GeV from Ref.~\cite{Moniz:1971mt}.

\section{Results and Discussions}

\begin{figure*}
\begin{minipage}[t]{0.48\textwidth}
\includegraphics[width=1.1\textwidth]{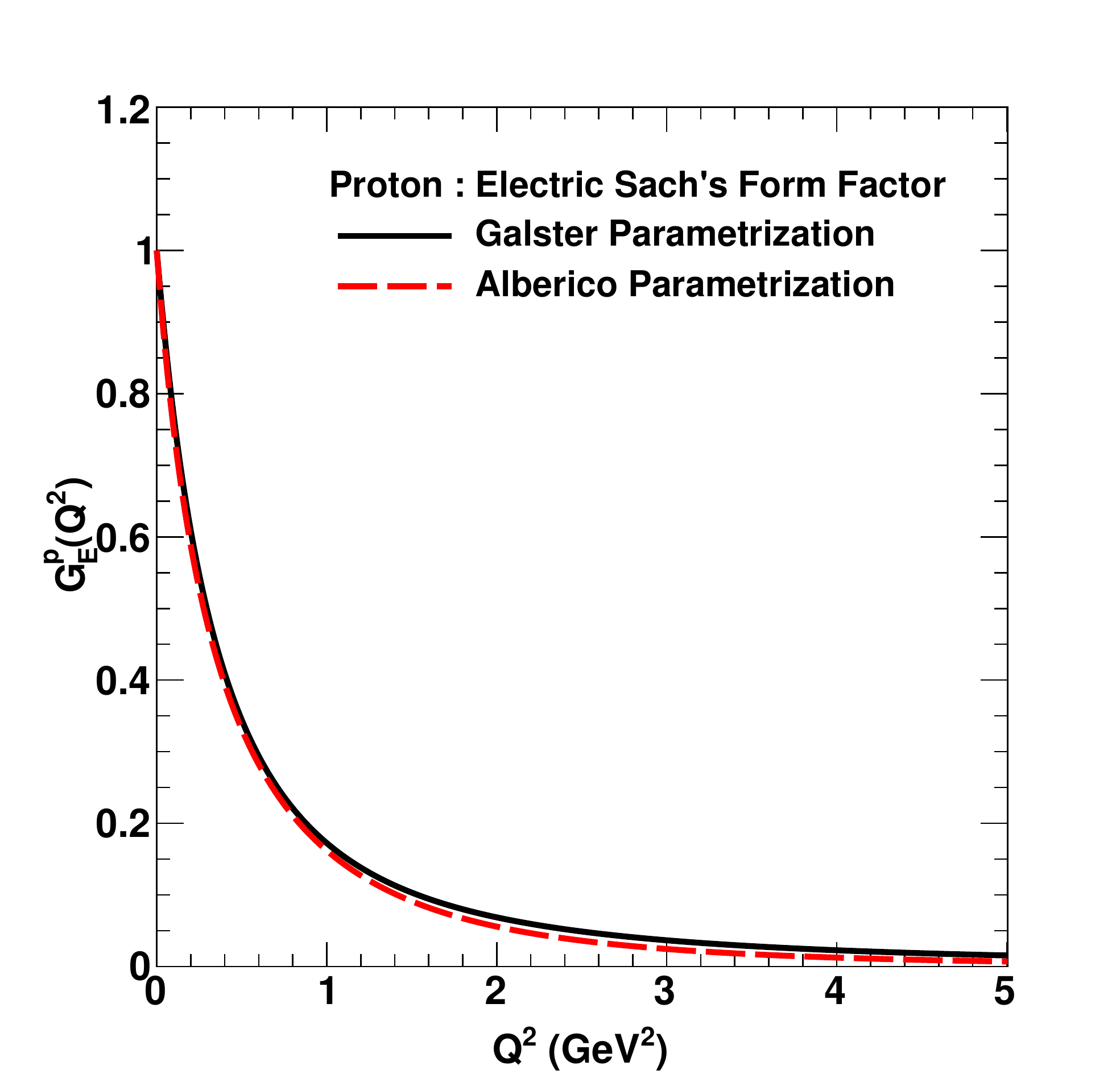}
\caption{Electric Sach's form factor $G^{p}_{E}$ for proton as a function of $Q^{2}$.}
\label{Figure1protonelectricsachdipole}
\end{minipage}
\hfill
\begin{minipage}[t]{0.48\textwidth}
\includegraphics[width=1.1\textwidth]{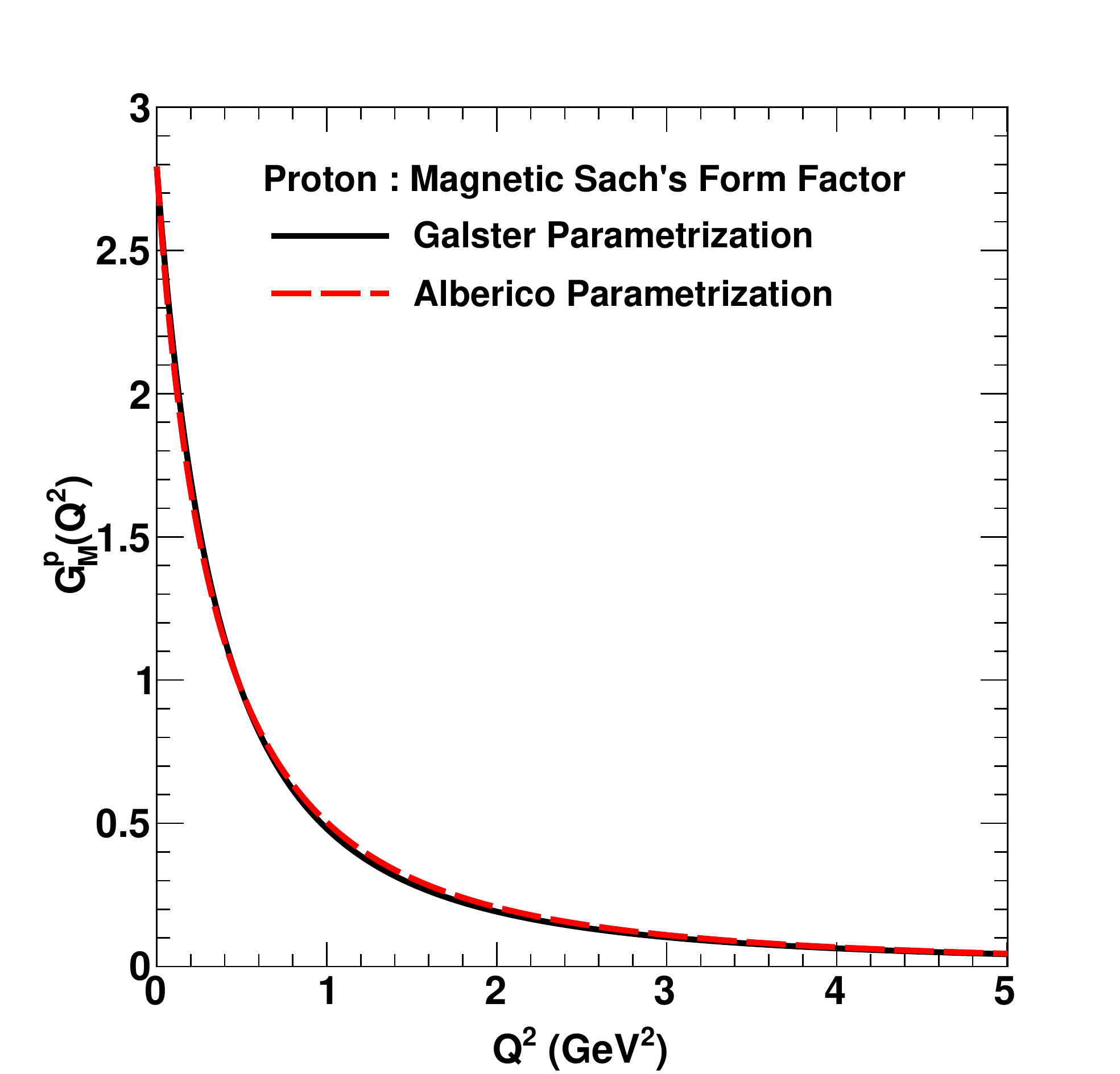}
\caption{Magnetic Sach's form factor $G^{p}_{M}$ for proton as a function of $Q^{2}$.}
\label{Figure2protonmagneticsachdipole}
\end{minipage}
\hfill
%\end{figure*}
%\begin{figure*}
\begin{minipage}[t]{0.48\textwidth}
\includegraphics[width=1.1\textwidth]{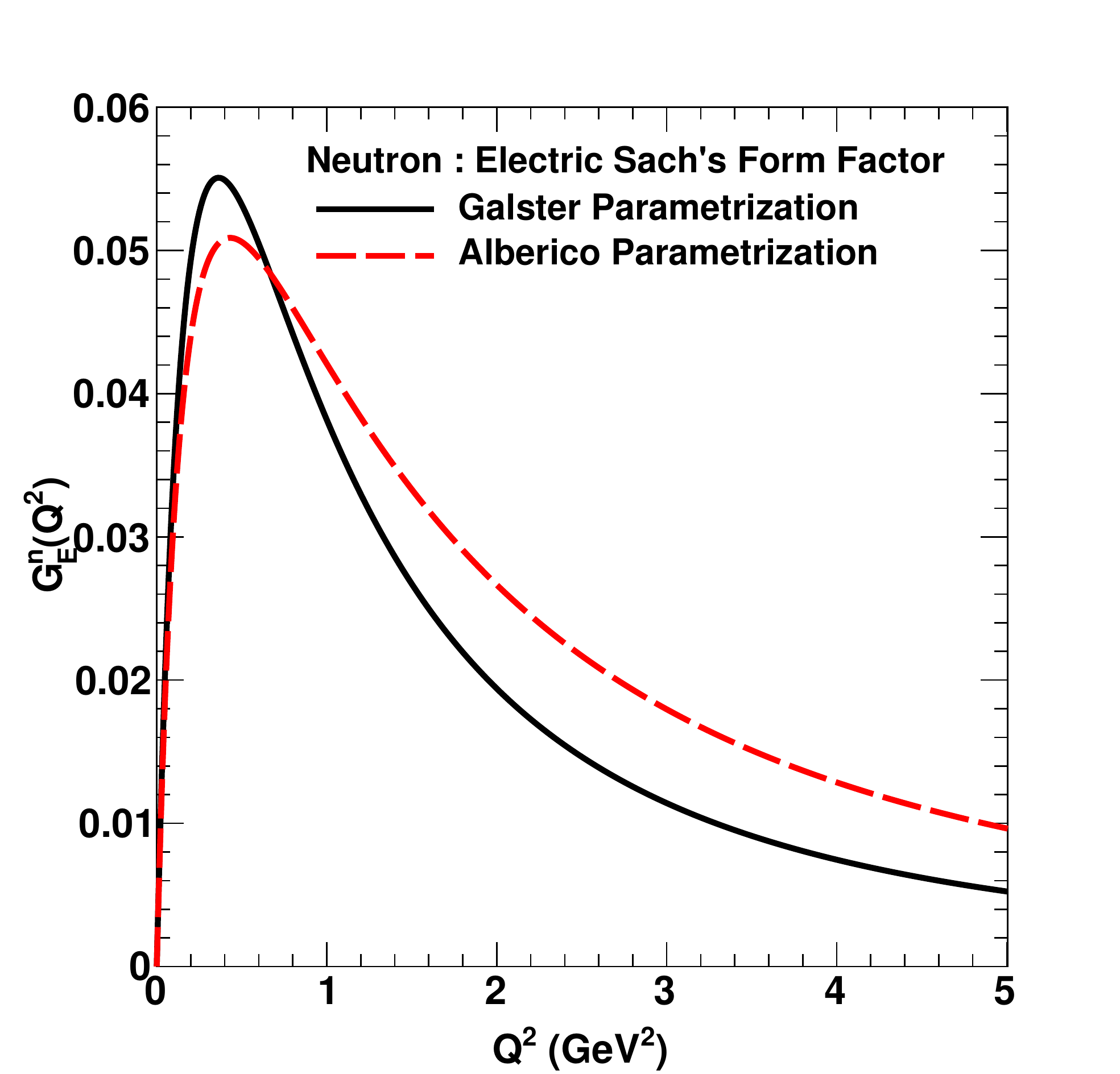}
\caption{Electric Sach's form factor for $G^{n}_{E}$ neutron as a function of $Q^{2}$.}
\label{Figure3neutronelectricsachdipole}
\end{minipage}
\hfill
\begin{minipage}[t]{0.48\textwidth}
\includegraphics[width=1.1\textwidth]{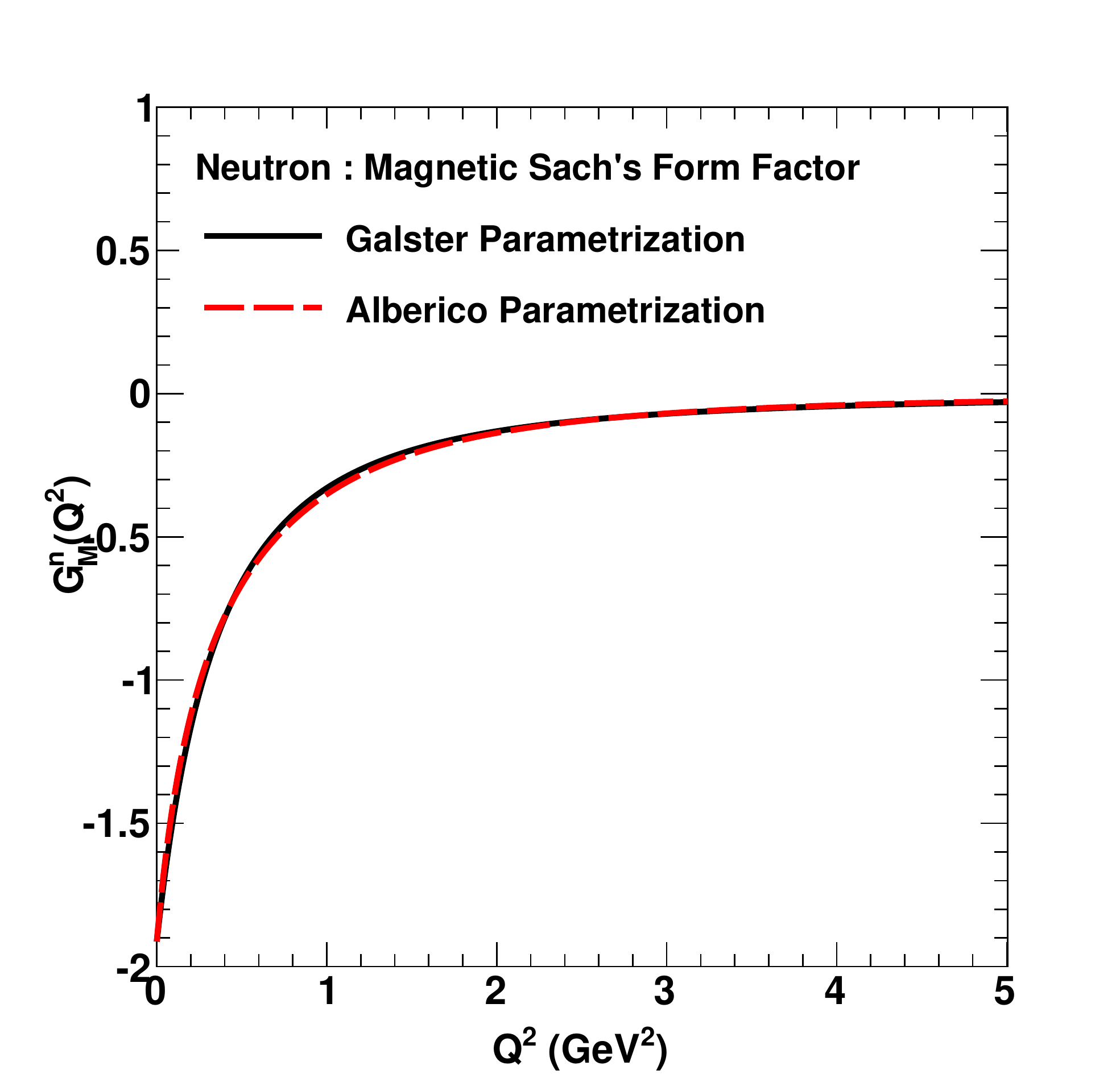}
\caption{Magnetic Sach's form factor for $G^{n}_{M}$ neutron as a function of $Q^{2}$.}
\label{Figure4neutronmagneticsachdipole}
\end{minipage}
\hfill
\end{figure*}

Figure \ref{Figure1protonelectricsachdipole} shows the electric Sach's 
form factor $G^{p}_{E}$ for proton as a function of square of momentum transfer 
$Q^{2}$ obtained using Galster and Alberico parametrizations. Both the parametrizations 
give very similar values of $G^{p}_{E}$ but at 
$Q^{2} \gtrsim$ 1 GeV$^{2}$ Galster's values are slightly above the Alberico's values.
Figure \ref{Figure2protonmagneticsachdipole} shows the magnetic Sach's 
form factor $G^{p}_{M}$ for proton  as a function of $Q^{2}$ obtained using Galster 
and Alberico parametrization. Both the parametrizations 
give very similar values of $G^{p}_{M}$ and for 
$Q^{2} \gtrsim$ 1 GeV$^{2}$ Galster's values are slightly below Alberico values.
Figure \ref{Figure3neutronelectricsachdipole} shows the electric Sach's 
form factor $G^{n}_{E}$ for neutron as a function of $Q^{2}$ obtained using 
Galster and Alberico parametrization. Below $Q^{2} \approx$ 1 GeV$^{2}$, 
the values of the Galster parametrization is more than that from Alberico 
parametrization but above $Q^{2} \approx$ 1 GeV$^{2}$, the Alberico parametrization 
gives higher values. Figure \ref{Figure4neutronmagneticsachdipole} shows the 
magnetic Sach's form factor $G^{n}_{M}$ for neutron as a function $Q^{2}$ obtained 
using the Galster and Alberico parametrization. Both the parametrizations give 
almost same values of $G^{n}_{M}$. 
In the Ref.~\cite{Akbar:2015yda}, the authors study various parameterizations 
of Galster et al.~\cite{Galster:1971kv}, Budd et al.~\cite{Budd:2004bp},
Bradford et al.~\cite{Bradford:2006yz},  Bosted et al.~\cite{Bosted:1994tm} 
and Alberico et al.~\cite{Alberico:2008sz} to observe their effect on scattering 
cross sections for neutrino induced CCQES processes on nuclear target like Ar.
They found that the cross section has no dependence on the choice of 
parameterizations at $E_{\nu} <$ 1 GeV.

\begin{figure*}
\begin{minipage}[t]{0.48\textwidth}
\includegraphics[width=1.1\textwidth]{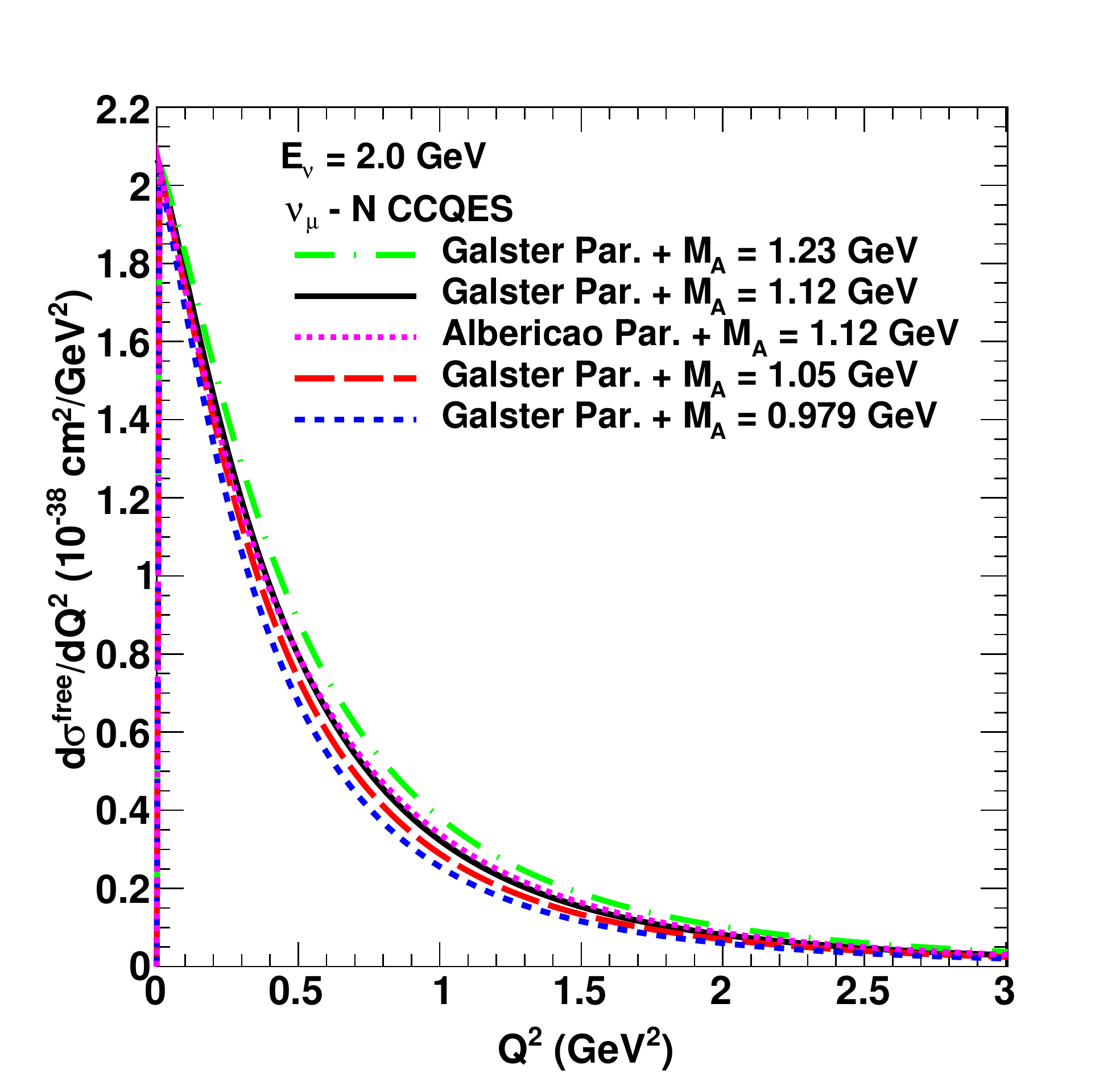}
\caption{Differential cross section $d\sigma/dQ^2$ for the neutrino-neutron 
CCQES as a function of $Q^2$ at neutrino energy $E_{\nu}$ = 2 GeV for different 
values of the axial mass $M_{A}$.}
\label{Figure5_neutrino_nucleon_differential_xsection_neutrino_two_gev}
\end{minipage}
\hfill
\begin{minipage}[t]{0.48\textwidth}
\includegraphics[width=1.1\textwidth]{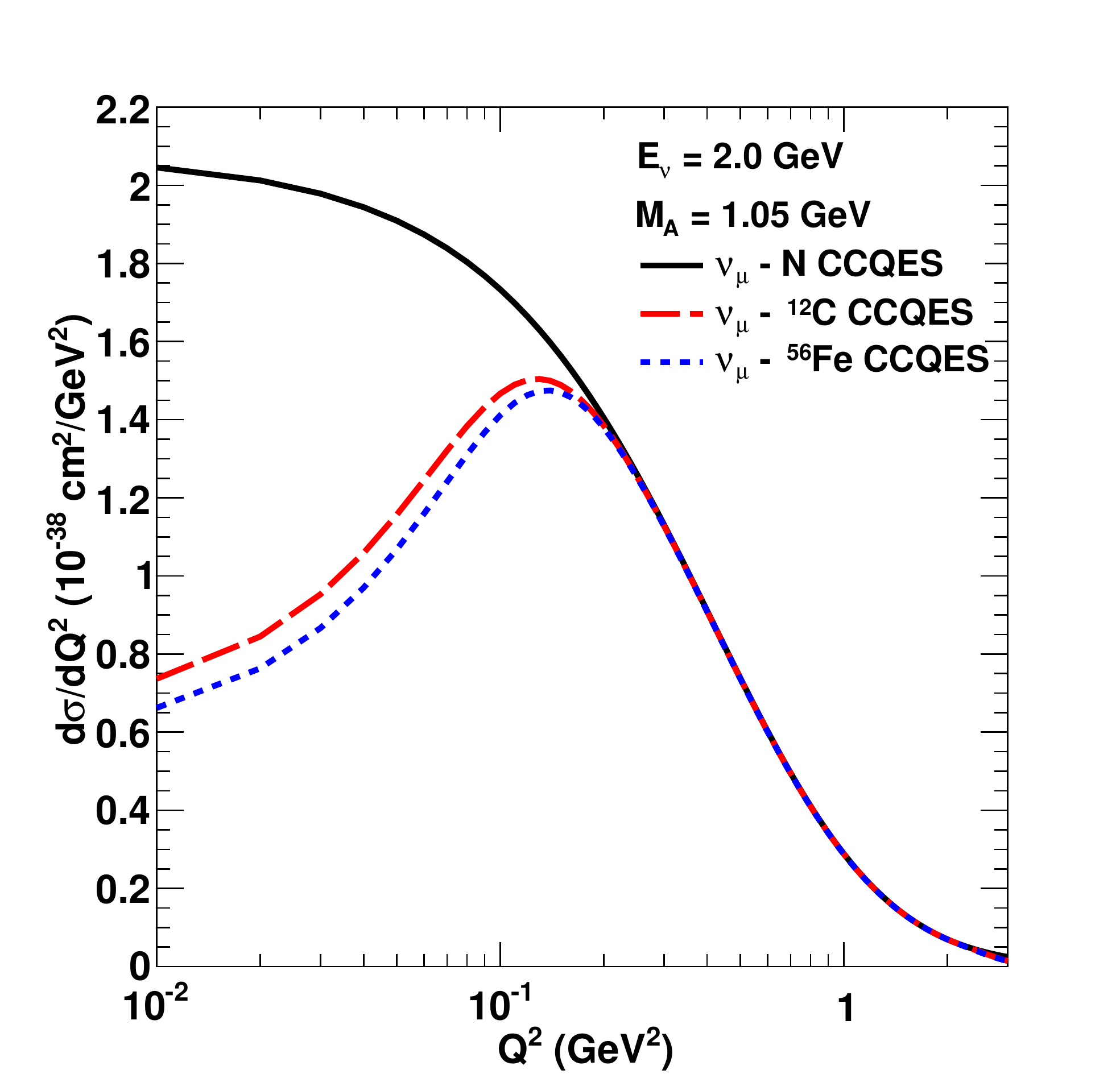}
\caption{Differential cross section $d\sigma/dQ^2$ for 
$\nu_{\mu}-N$, $\nu_{\mu}-^{12}C$ and $\nu_{\mu}-^{56}Fe$ CCQES as a function of $Q^2$ 
at neutrino energy $E_{\nu}$ = 2 GeV for axial mass $M_{A}$ = 1.05 GeV.}
\label{Figure6_neutrino_nucleon_carbon_iron_differential_xsection_neutrino_two_gev}
\end{minipage}
\hfill
\end{figure*}

Figure \ref{Figure5_neutrino_nucleon_differential_xsection_neutrino_two_gev} 
shows the differential cross section $d\sigma/dQ^2$ for the neutrino-neutron 
CCQES as a function of $Q^2$ obtained using Galster (at $M_{A}$= 0.979, 1.05, 
1.12 and 1.23 GeV) and Alberico (at $M_{A}$= 1.12 GeV) parametrization at 2 
GeV neutrino energy. The value of $d\sigma/dQ^2$ increases with the increase 
in the value of axial mass $M_{A}$. A comparison of the cross sections using the 
Alberico and the Galster parametrizations made at axial mass $M_{A}$ = 1.12 GeV 
shows that there is no noticeable difference due to different parametrizations 
of Sach's form factors and for all further calculations we use the Galster 
parametrization. Such a study has been made with all the parameterizations as mentioned
in the last section and these do not make any difference to the cross 
section.

Figure \ref{Figure6_neutrino_nucleon_carbon_iron_differential_xsection_neutrino_two_gev}  
shows the differential cross section $d\sigma/dQ^2$ for the neutrino-neutron,
neutrino-carbon and neutrino-iron CCQES as a function of $Q^2$ at 2 GeV 
neutrino energy obtained using axial mass $M_{A}$= 1.05 GeV. For the neutrino-carbon
and neutrino-iron calculations we use the Fermi Gas model with Pauli 
blocking. The Fermi momentum ($k_{F}$) for the carbon nucleus is taken as 0.221 
GeV and for the iron nucleus it is taken as 0.260 GeV. The effective binding energy 
$E_B$ of nucleon in the nucleus is taken as 10 MeV. The value of $a$ is 0.020 GeV. 
The cross sections at low $Q^2$ drop due to the nuclear effects.
Due to diffuseness parameter, the differential cross-section for neutrino nucleus 
interaction drops smoothly as $Q^{2}$ goes to zero as shown in Fig. 
\ref{Figure6_neutrino_nucleon_carbon_iron_differential_xsection_neutrino_two_gev}.

\begin{figure*}
\begin{minipage}[t]{0.48\textwidth}
\includegraphics[width=1.1\textwidth]{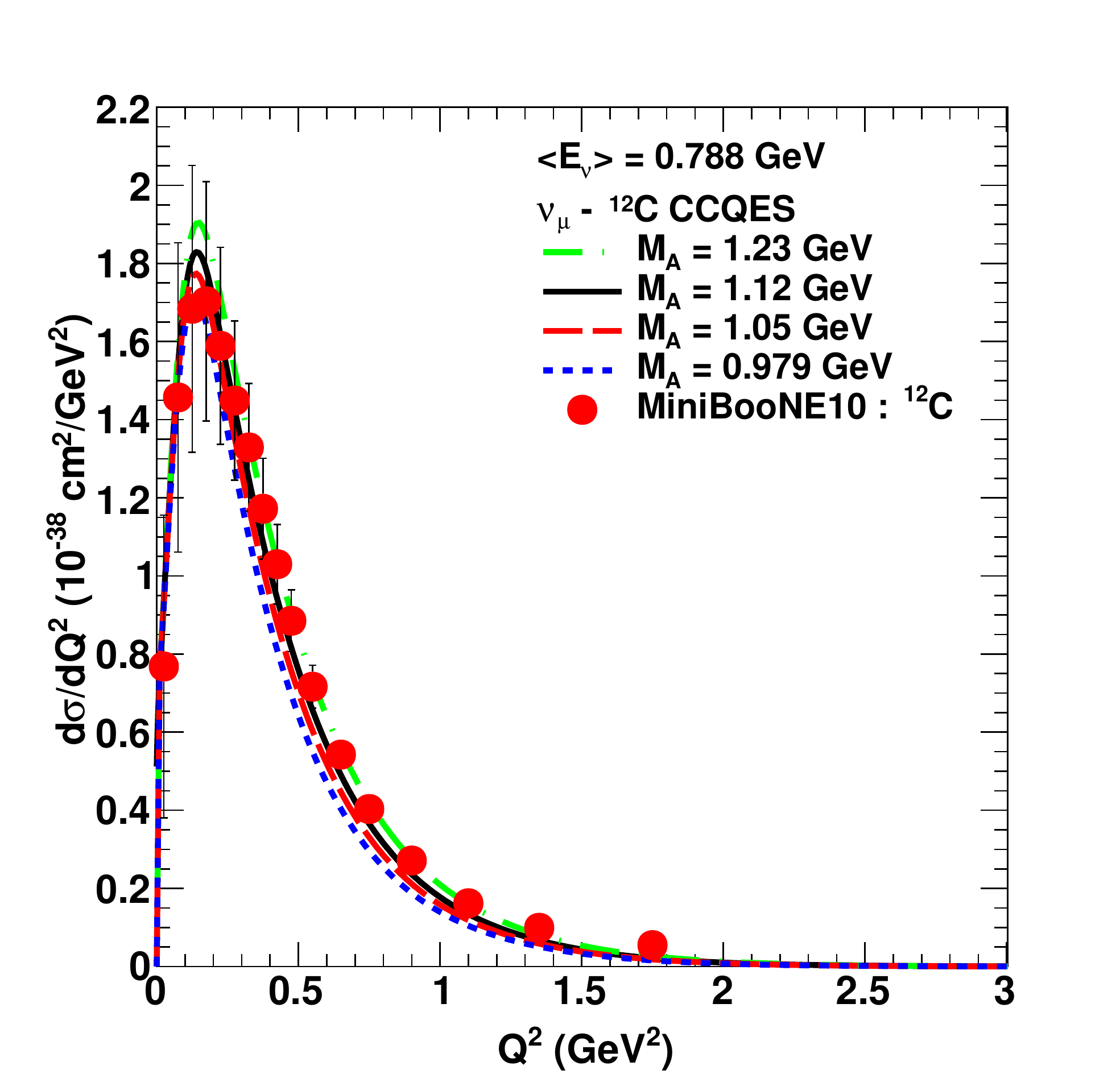}
\caption{Differential cross section $d\sigma/dQ^2$ per neutron for 
the neutrino-carbon CCQES as a function of $Q^2$ for different values of the 
axial mass $M_{A}$ at an average neutrino energy $<E_{\nu}>$ = 0.788 GeV 
corresponding to MiniBooNE data \cite{AguilarArevalo:2010zc}.}
\label{Figure7_neutrino_carbon_differential_xsection_neutrino_788_mev}
\end{minipage}
\hfill
\begin{minipage}[t]{0.48\textwidth}
\includegraphics[width=1.1\textwidth]{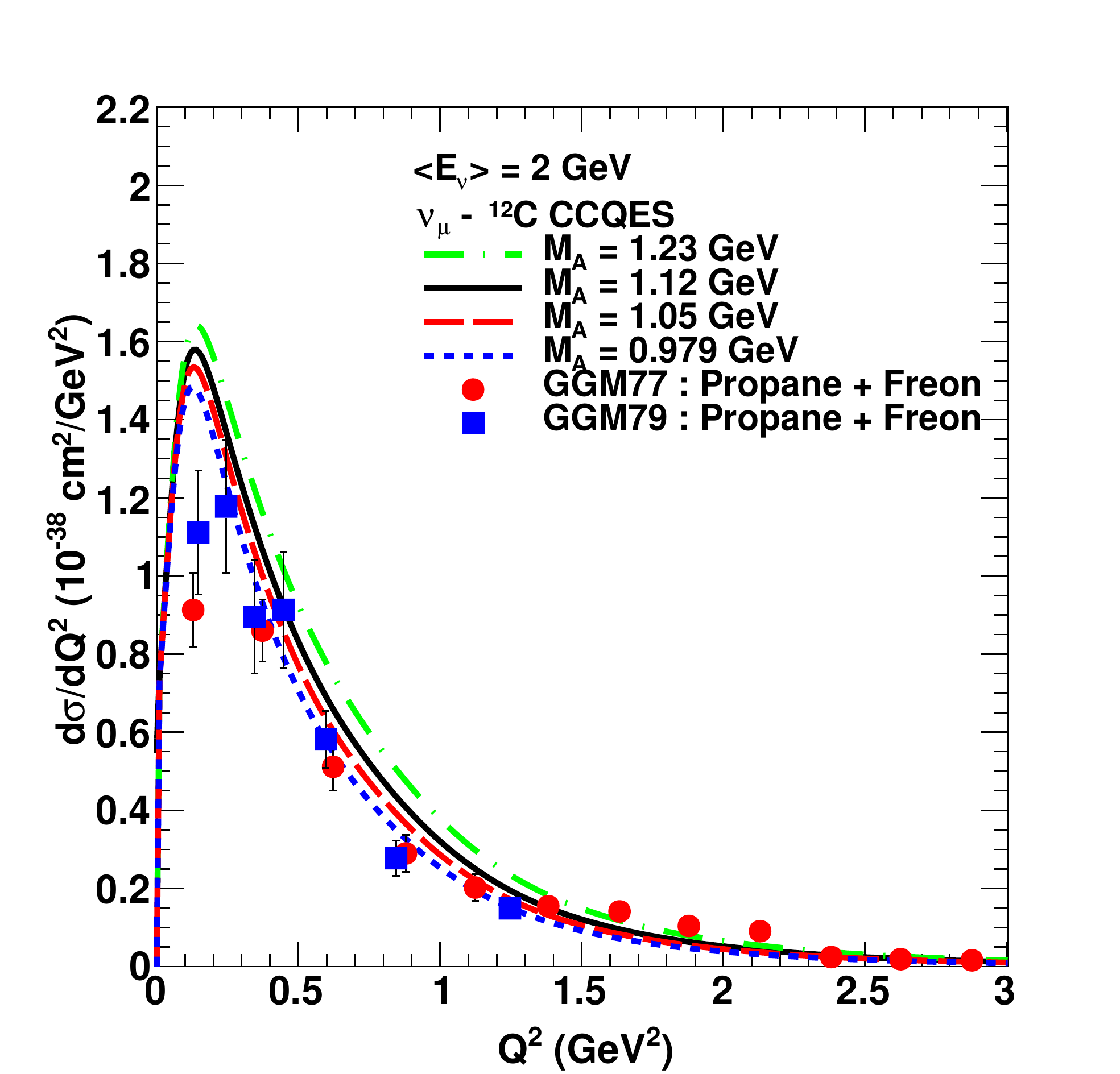}
\caption{Differential cross section $d\sigma/dQ^2$ per neutron for the neutrino-carbon
  CCQES as a function of $Q^2$ for different values of the axial mass 
$M_{A}$ at average neutrino energy $<E_{\nu}>$ = 2 GeV corresponding to GGM data 
\cite{Bonetti:1977cs, Pohl:1979zm}.}
\label{Figure8_neutrino_carbon_differential_xsection_neutrino_two_gev}
\end{minipage}
\hfill
\end{figure*}

Figure \ref{Figure7_neutrino_carbon_differential_xsection_neutrino_788_mev} shows 
the differential cross section $d\sigma/dQ^2$ per neutron for the neutrino-carbon 
CCQES as a function of $Q^2$ at different values of axial mass $M_{A}$= 0.979, 
1.05, 1.12 and 1.23 GeV using Galster parametrizations for Sach's form factors.
  The calculations correspond to an average neutrino energy $<E_{\nu}>$ = 0.788 
GeV are compared with the data recorded by the MiniBooNE (MiniBooNE10) experiment 
\cite{AguilarArevalo:2010zc}. The cross section is obtained by averaging over 
calculations in a range of energies weighted by the neutrino energy spectrum  
given by the MiniBooNE data. The calculations with $M_{A}$= 1.05, 1.12 and 1.23 
GeV are compatible with the data.

 Figure \ref{Figure8_neutrino_carbon_differential_xsection_neutrino_two_gev} 
shows the differential cross section $d\sigma/dQ^2$ per neutron for the 
neutrino-carbon CCQES as a function of $Q^2$ with different values of 
$M_{A}$= 0.979, 1.05, 1.12 and 1.23 GeV.  The calculations are at an average 
neutrino energy $<E_{\nu}>$ = 2 GeV corresponding to the data recorded by the 
GGM77 (Gargamelle) \cite{Bonetti:1977cs}.  The cross section from GGM79 
 experiment \cite{Pohl:1979zm} measured in the neutrino energy 
range 1.5 to 5.5 GeV are also plotted. The calculations with $M_{A}$= 0.979 
and 1.05 GeV are compatible with the data.

\begin{figure*}
\begin{minipage}[t]{0.48\textwidth}
\includegraphics[width=1.1\textwidth]{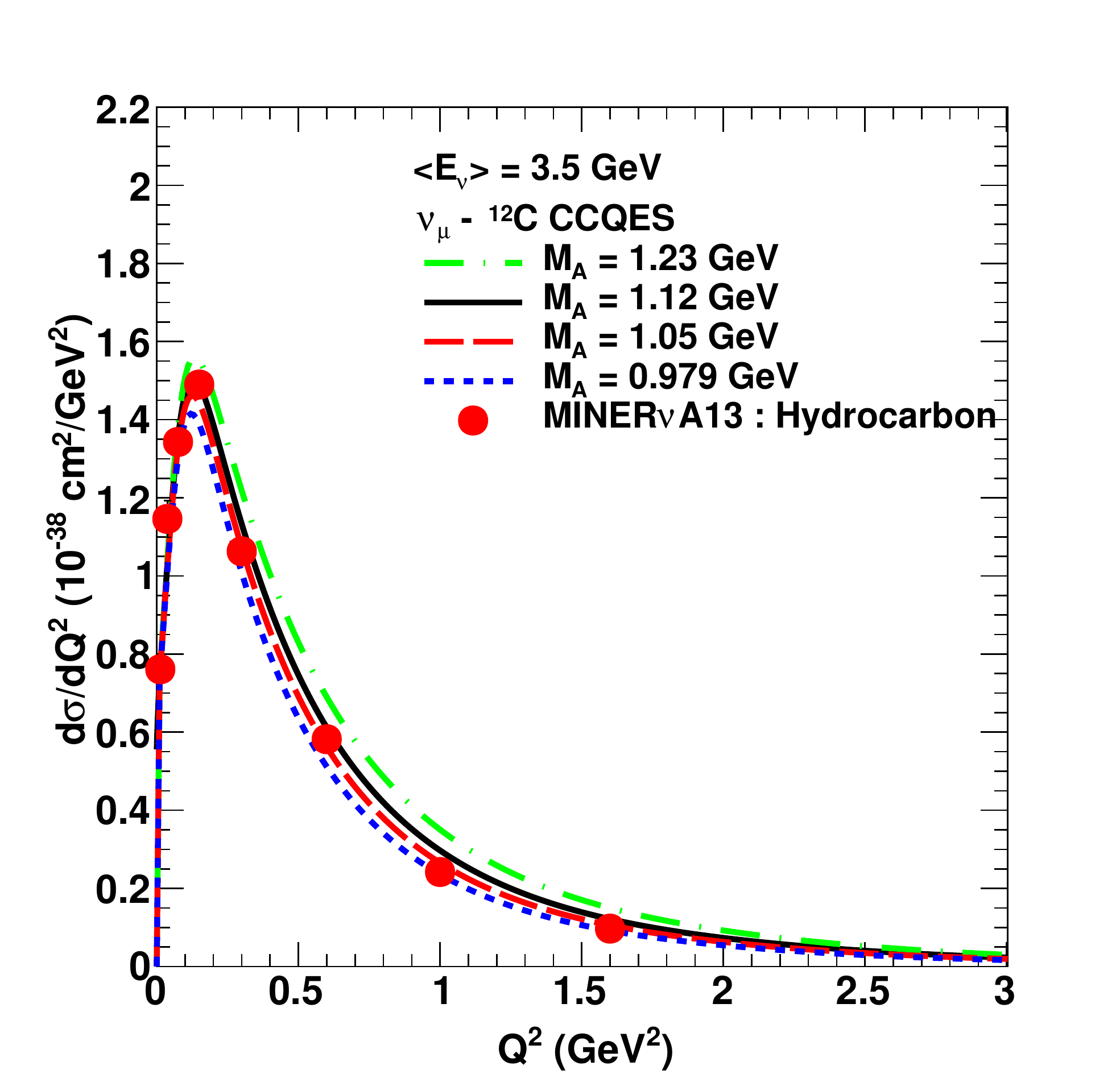}
\caption{Differential cross section $d\sigma/dQ^2$ per neutron for the neutrino-carbon
  CCQES as a function of $Q^2$ for different values of the axial mass 
$M_{A}$ at average neutrino energy $<E_{\nu}>$ = 3.5 GeV  corresponding to 
MINER$\nu$A data \cite{Fiorentini:2013ezn}.}
\label{Figure9_neutrino_carbon_differential_xsection_neutrino_35_gev}
\end{minipage}
\hfill
\begin{minipage}[t]{0.48\textwidth}
\includegraphics[width=1.1\textwidth]{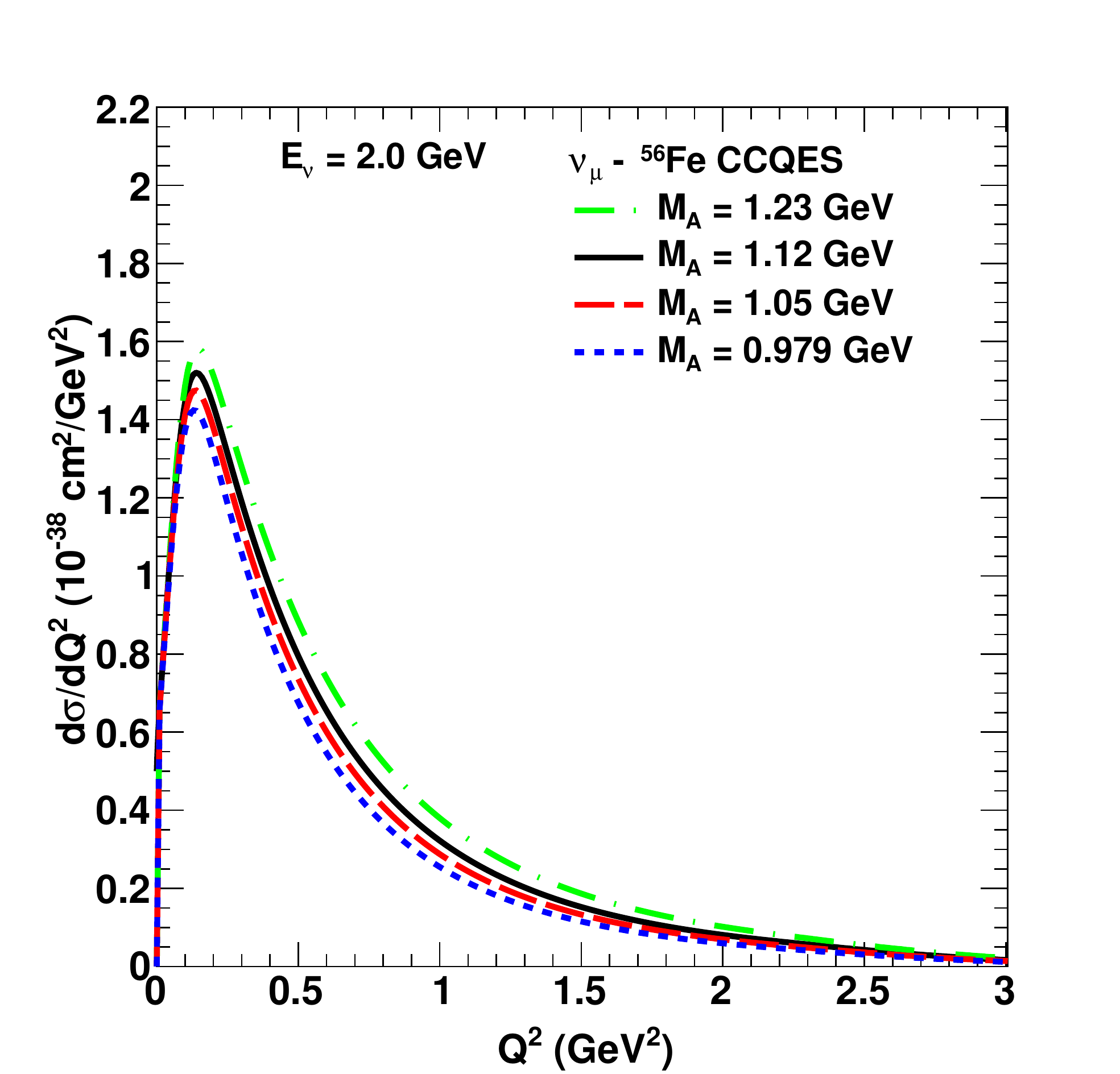}
\caption{Differential cross section $d\sigma/dQ^2$ per neutron for the neutrino-iron
  CCQES as a function of $Q^2$ at neutrino energy $E_{\nu}$ = 2 GeV for different 
values of the axial mass $M_{A}$.}
\label{Figure10_neutrino_iron_differential_xsection_neutrino_two_gev}
\end{minipage}
\hfill
\end{figure*}

Figure \ref{Figure9_neutrino_carbon_differential_xsection_neutrino_35_gev} shows 
the differential cross section $d\sigma/dQ^2$ per neutron for the neutrino-carbon 
CCQES as a function of $Q^2$ with values of $M_{A}$= 0.979, 1.05, 1.12 and 
1.23 GeV.  The calculations are at average neutrino energy $<E_{\nu}>$ = 3.5 GeV 
corresponding to the MINER$\nu$A data \cite{Fiorentini:2013ezn}. The calculations 
with $M_{A}$= 1.05 and 1.12 GeV are compatible with the data.

 Figure \ref{Figure10_neutrino_iron_differential_xsection_neutrino_two_gev} 
predicts the differential cross section $d\sigma/dQ^2$ per neutron for the 
neutrino-iron CCQES as a function of the square of momentum transfer $Q^2$ at 
neutrino energy 2 GeV for values of $M_{A}$= 0.979, 1.05, 1.12 and 1.23 GeV.

\begin{figure*}
\begin{minipage}[t]{0.48\textwidth}
\includegraphics[width=1.1\textwidth]{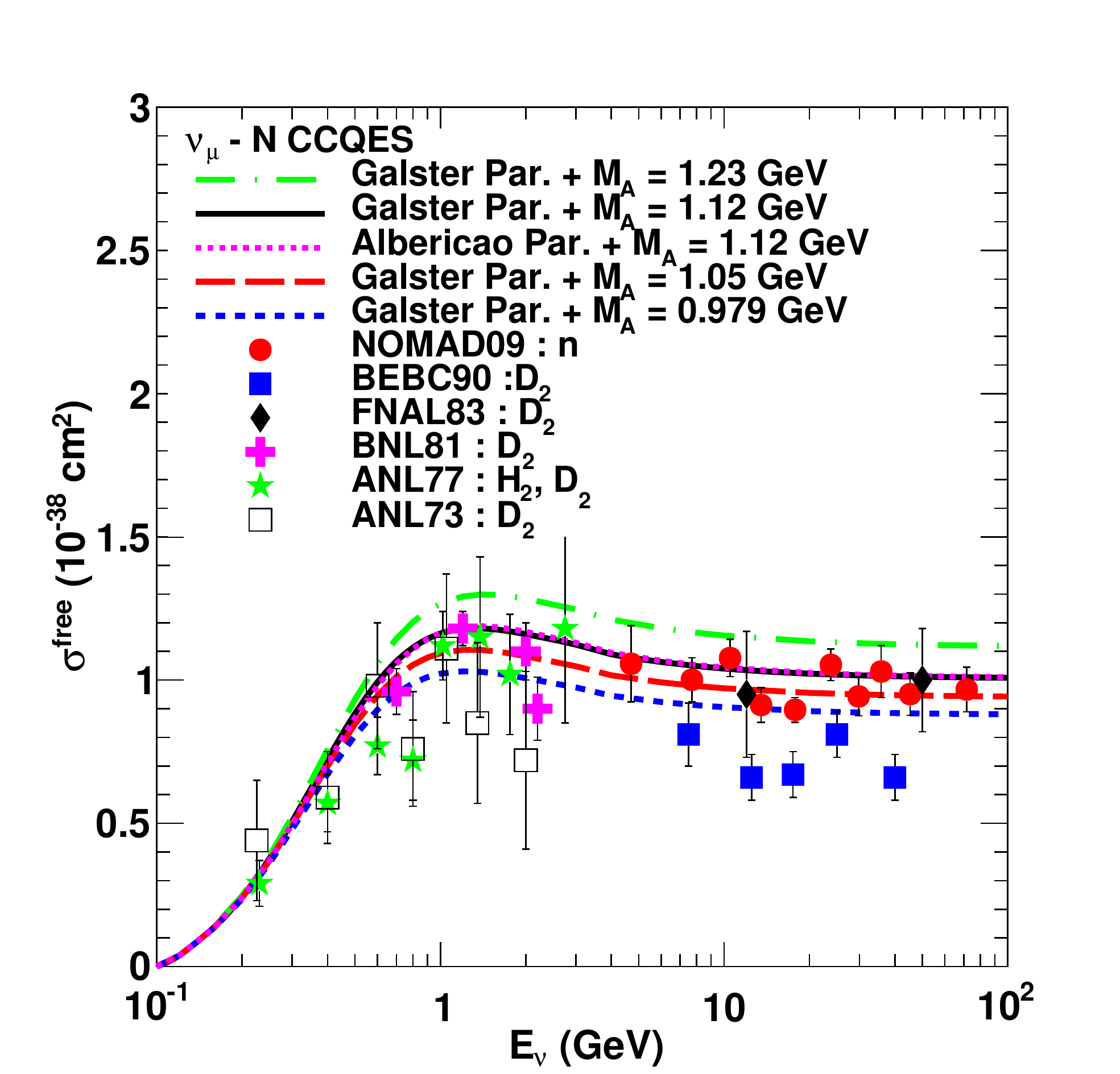}
\caption{Total cross section $\sigma$ for the neutrino-neutron CCQES as a 
function of neutrino energy $E_{\nu}$ for different values of the axial mass 
$M_{A}$ compared with the data 
\cite{Lyubushkin:2008pe, Mann:1973pr, Barish:1977qk, Baker:1981su, Kitagaki:1983px, Allasia:1990uy}}
\label{Figure11_neutrino_nucleon_total_cross_section}
\end{minipage}
\hfill
\begin{minipage}[t]{0.48\textwidth}
\includegraphics[width=1.1\textwidth]{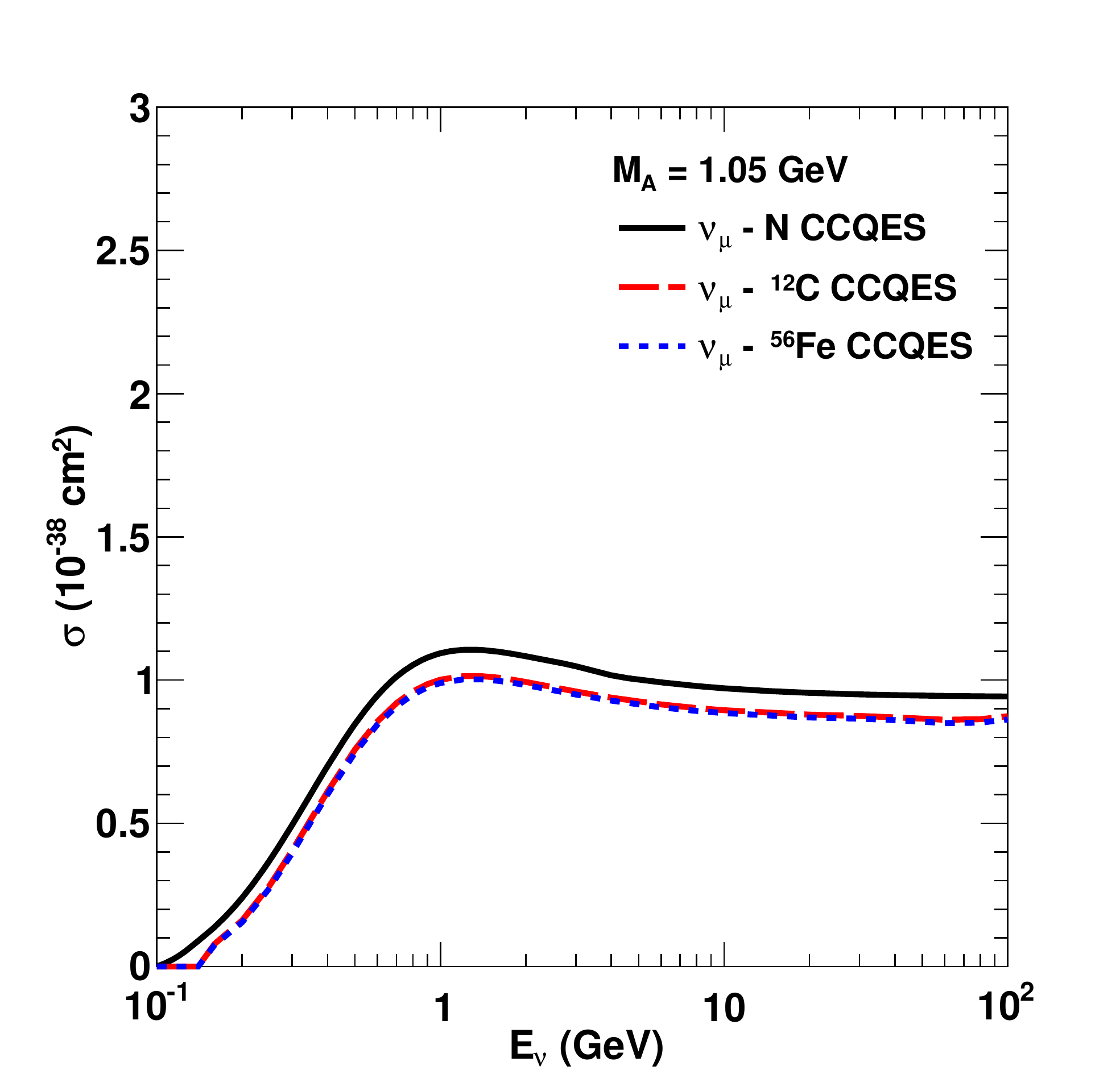}
\caption{Total cross section $\sigma$ for $\nu_{\mu}-N$, $\nu_{\mu}- ^{12}C$ and 
$\nu_{\mu} - ^{56}Fe$ CCQES as a function of neutrino energy $E_{\nu}$ for axial 
mass $M_{A}$ = 1.05 GeV.}
\label{Figure12_neutrino_nucleon_carbon_iron_total_xsection}
\end{minipage}
\hfill
\end{figure*}

Figure \ref{Figure11_neutrino_nucleon_total_cross_section} shows the total cross 
section $\sigma$ for the neutrino - neutron CCQES as a function of $E_{\nu}$ 
obtained using Galster (at $M_{A}$= 0.979, 1.05, 1.12 and 1.23 GeV) and Alberico 
(at $M_{A}$= 1.12 GeV) parametrization. The value of $\sigma$ increases with the 
increase in the value of axial mass $M_{A}$. A comparison of the total cross 
sections using the Alberico parametrizations and the Galster parametrization 
made at axial mass $M_{A}$ = 1.12 GeV shows that there is no noticeable 
difference due to different parametrizations of Sach's form factors.
  The calculations are compared with the data recorded by 
Neutrino Oscillation MAgnetic Detector (NOMAD09) \cite{Lyubushkin:2008pe}, 
Argonne National Laboratory (ANL73) \cite{Mann:1973pr},  
Argonne National Laboratory (ANL77) \cite{Barish:1977qk}, 
Brookhaven National Laboratory (BNL81) \cite{Baker:1981su}, 
Fermi National Laboratory (FNAL83) \cite{Kitagaki:1983px} and  
Big European Bubble Chamber (BEBC90) \cite{Allasia:1990uy} collaboration. 
The calculations with $M_{A}$= 0.979, 1.05 and 1.12 GeV are 
compatible with the data. The calculations of $\sigma$ with $M_{A}$ = 1.23 GeV 
overestimate all the experimental data. The diffuseness parameter does not affect the 
total cross section.

Figure \ref{Figure12_neutrino_nucleon_carbon_iron_total_xsection} shows the 
total cross section $\sigma$ for the neutrino-neutron, neutrino-carbon and 
neutrino-iron CCQES as a function of $E_{\nu}$ obtained using axial mass 
$M_{A}$= 1.05 GeV.  For the neutrino-carbon and neutrino-iron calculations 
we use the Fermi Gas model with Pauli blocking. The nuclear effects reduce 
the cross section by 10 \% even at higher neutrino energy above 1 GeV.
%The value of $E_B$ is reduced to 10 MeV which gives good description of the 
%neutrino-iron data at low energies.

\begin{figure*}
\begin{minipage}[t]{0.48\textwidth}
\includegraphics[width=1.1\textwidth]{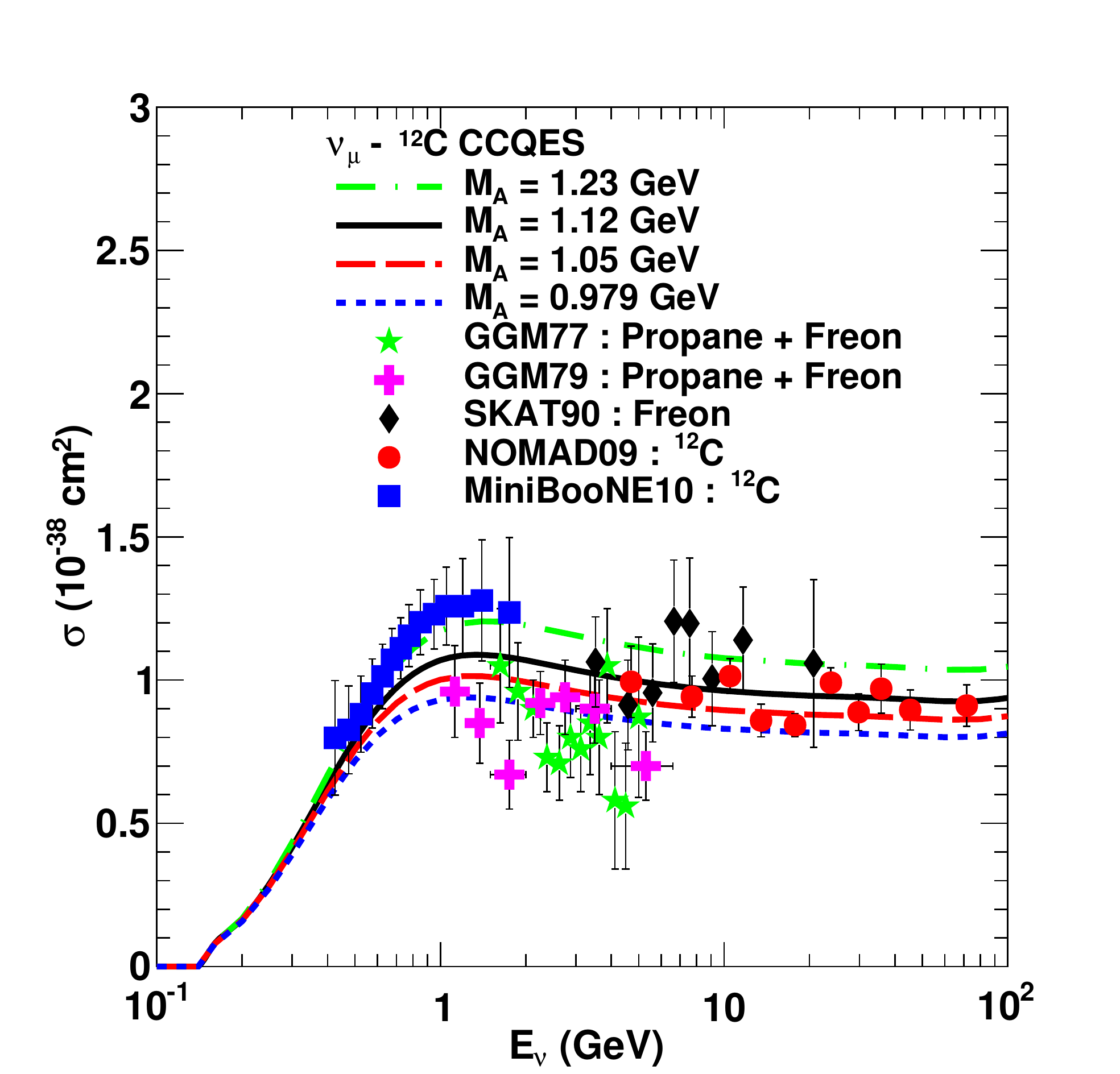}
\caption{Total cross section $\sigma$ per neutron for the neutrino-carbon 
charged current quasi elastic scattering as a function of neutrino energy 
$E_{\nu}$ for different values of the axial mass $M_{A}$ compared with the data 
\cite{Lyubushkin:2008pe, AguilarArevalo:2010zc, Bonetti:1977cs, Pohl:1979zm, Brunner:1989kw}.}
\label{Figure13_neutrino_carbon_total_cross_section}
\end{minipage}
\hfill
\begin{minipage}[t]{0.48\textwidth}
\includegraphics[width=1.1\textwidth]{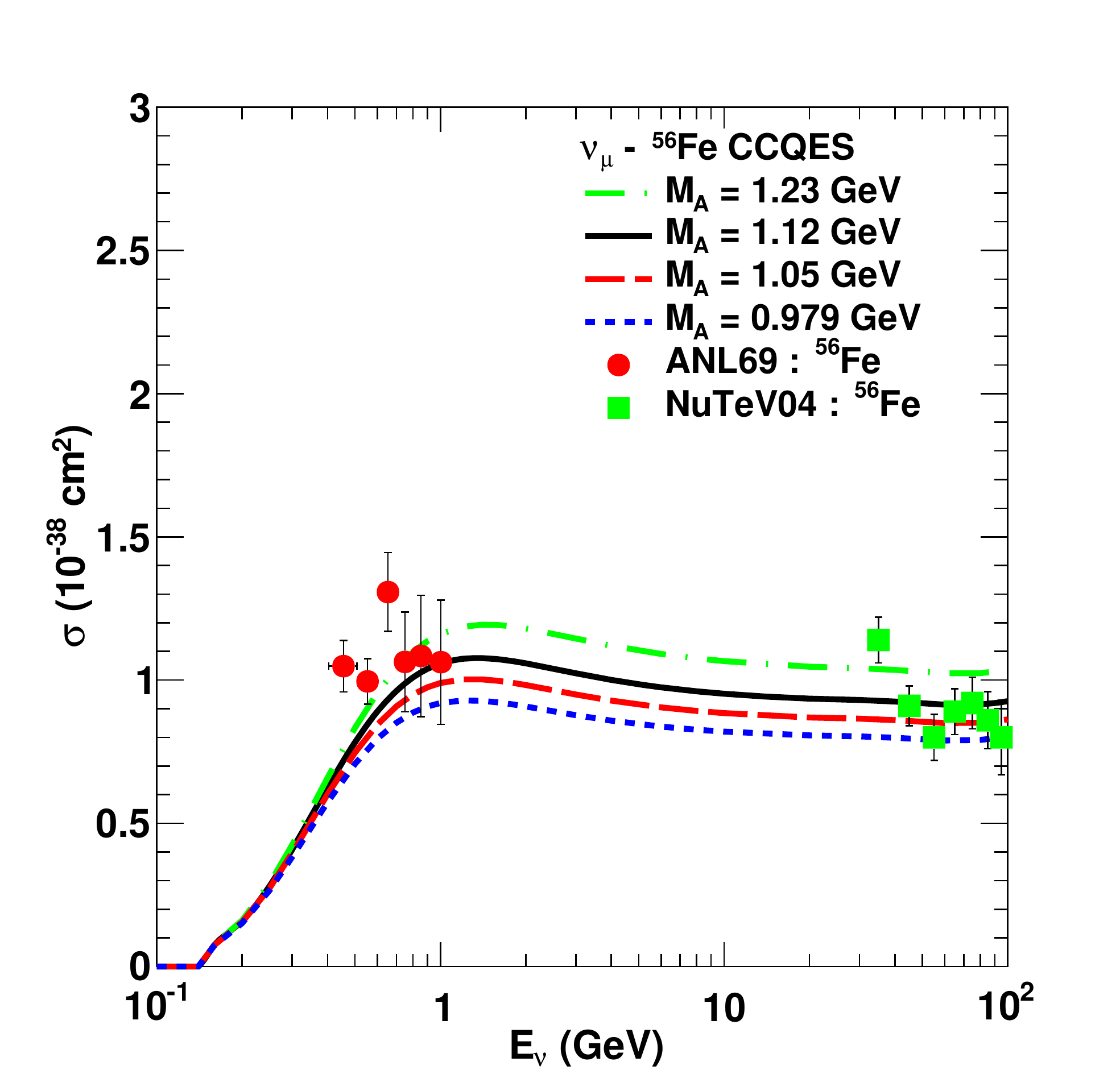}
\caption{Total cross section $\sigma$ per neutron for the neutrino-iron charged 
current quasi elastic scattering as a function of neutrino energy $E_{\nu}$ for 
different values of the axial mass $M_{A}$ compared with the data 
\cite{Kustom:1969dh, Suwonjandee:2004aw}.}
\label{Figure14_neutrino_iron_total_cross_section}
\end{minipage}
\hfill
\end{figure*}

Figure \ref{Figure13_neutrino_carbon_total_cross_section} shows the total cross 
section $\sigma$ per neutron for the neutrino-carbon CCQES scattering as a 
function of $E_{\nu}$ obtained using Galster parametrization with axial mass 
$M_{A}$= 0.979, 1.05, 1.12 and 1.23 GeV. The calculations are 
compared with the data recorded by NOMAD09 \cite{Lyubushkin:2008pe}, 
MiniBooNE10 \cite{AguilarArevalo:2010zc}, GGM77 \cite{Bonetti:1977cs}, 
GGM79 \cite{Pohl:1979zm} and SKAT90 \cite{Brunner:1989kw} experiments. The 
calculations with the value of $M_{A}$= 1.23 GeV describes the MiniBooNE data but overestimate
the other experimental data. The calculations with $M_{A}$= 1.05 and 1.12 GeV 
are compatible with NOMAD09, GGM77, GGM79 and SKAT90 data. 
There are alternative ways to get a better 
description of the cross-section data at low energy e.g. the work in 
Ref.~\cite{Kolupaeva:2016bfg} parametrizes $M_A$ as a function of energy which 
results in higher value of $M_A$ at low energy.

Figure \ref{Figure14_neutrino_iron_total_cross_section} shows the total cross 
section $\sigma$ per neutron for the neutrino-iron CCQES as a function of the 
$E_{\nu}$ obtained using Galster parametrization with axial mass $M_{A}$= 0.979, 
1.05, 1.12 and 1.23 GeV. The calculations are compared with the data recorded by 
ANL69 \cite{Kustom:1969dh} and Neutrino at the Tevatron 
(NuTeV04) \cite{Suwonjandee:2004aw} experiments. The calculations with 
$M_{A}$= 1.05 and 1.12 GeV are compatible with the data.

\section{Conclusion}

We presented a study on the charge current quasi elastic scattering of $\nu_\mu$ 
from nucleon and nuclei. We use a Fermi model with Pauli suppression factor which is 
simple to incorporate yet includes all the essential features. 
The investigation of parametrizations for 
electric and magnetic Sach's form factors of nucleons shows that there is 
no noticeable difference in cross section due to different parametrizations.
  Calculations have been made for CCQES total and differential cross sections
for the cases of $\nu_{\mu}-N$, $\nu_{\mu}-^{12}C$ and $\nu_{\mu}-^{56}Fe$ scatterings
and are compared with the data for different values of the axial mass.
The calculations give excellent description of the differential cross section 
data. 
%The Pauli Suppression effects are found to reduce the cross section by 10 \% even at higher 
%neutrino energy above 1 GeV. 
The diffuseness parameter does not affect the total cross section. 
The calculations with axial mass 1.05 and 1.12 GeV give 
good description of most of the experimental data and thus a value between these 
two can be taken as the most acceptable value of $M_A$.  The data from MiniBooNE 
demands a larger value of $M_{A}$= 1.23 GeV to get an excellent fit.

\end{document}